\documentclass[reprint,showpacs,showkeys,amsmath,amssymb,twocolumn,aps,superscriptaddress,floatfix]{revtex4-2}
\usepackage{amssymb} 
\usepackage{amsmath,bm} 
\usepackage{graphicx} 
\usepackage[normalem]{ulem} 
\usepackage{multirow} 
\usepackage[colorlinks,linkcolor=blue,urlcolor=blue,citecolor=blue]{hyperref} 
\usepackage{lipsum} 
\usepackage[usenames, dvipsnames]{xcolor} 
\usepackage{tensor} 
\usepackage{isotope} 
\usepackage{amsmath} 
\usepackage{url} 
\allowdisplaybreaks[4] 

\renewcommand{\sout}{\bgroup \color{red} \ULdepth=-.5ex \ULset}

\begin{document}

\title{Atmospheric Antideuteron Flux  Within a Dynamical Coalescence Approach
}
\author{Jie Pu}
\email{pujie@htu.edu.cn}
\affiliation{College of Physics, Centre for Theoretical Physics, Henan Normal University, Xinxiang 453007, China}
\affiliation{Shanghai Research Center for Theoretical Nuclear Physics, NSFC and Fudan University, Shanghai 200438, China}

\author{Xin Li}
\affiliation{College of Physics, Centre for Theoretical Physics, Henan Normal University, Xinxiang 453007, China}
\affiliation{State Key Laboratory of Dark Matter Physics, Key Laboratory for Particle Astrophysics and Cosmology (MOE), and Shanghai Key Laboratory for Particle Physics and Cosmology, School of Physics and Astronomy, Shanghai Jiao Tong University, Shanghai 200240, China}

\author{Kai-Jia Sun}
\email{Corresponding author: kjsun$@$fudan.edu.cn}
\affiliation{Key Laboratory of Nuclear Physics and Ion-beam Application (MOE), Institute of Modern Physics,  Fudan University, Shanghai 200433, China}
\affiliation{Shanghai Research Center for Theoretical Nuclear Physics, NSFC and Fudan University, Shanghai 200438, China}
\author{Chun-Wang Ma}
\email{Corresponding author: machunwang$@$126.com}
\affiliation{Institute of Nuclear Science and Technology, Henan Academy of Sciences, Zhengzhou 450046, China}
\affiliation{College of Physics, Centre for Theoretical Physics, Henan Normal University, Xinxiang 453007, China}

\author{Lie-Wen Chen}
\email{Corresponding author: lwchen$@$sjtu.edu.cn}
\affiliation{State Key Laboratory of Dark Matter Physics, Key Laboratory for Particle Astrophysics and Cosmology (MOE), and Shanghai Key Laboratory for Particle Physics and Cosmology, School of Physics and Astronomy, Shanghai Jiao Tong University, Shanghai 200240, China}

\date{\today}
   
\begin{abstract}
Cosmic antideuterons are considered as one of the most promising tools for the indirect detection of dark matter due to their ultra-low astrophysical backgrounds. Currently only upper limits on the antideuteron flux exist, but advancements in experimental detection technology may soon lead to positive signals. A major source of background is the production of secondary antideuterons through collisions of cosmic rays with the surrounding medium. In this study, antideuteron production is modeled using a multiphase transport model (AMPT) coupled with a dynamical coalescence model. By applying a widely used leaky box model and incorporating specific processes, we present a new theoretical baseline for atmospheric secondary antideuteron flux, including a tertiary contribution, from primary cosmic rays interacting with Earth's atmosphere. Our results indicate that the atmospheric antideuteron flux are within the range of various existing calculations and remain well below the upper limits set by the Balloon-borne Experiment with a Superconducting Spectrometer (BESS). The atmospheric antideuteron is found to dominate the antideuteron background at kinetic energies below $0.26 $ GeV/n.
\end{abstract}
\maketitle

\section{introduction}
Searching for cosmic antideuterons has been proposed as a promising method to indirectly detect dark matter (DM) annihilations and decays~\cite{Donato:1999gy,Baer:2005tw}, due to their ultra-low astrophysical backgrounds. The production of secondary antideuterons at low energies is significantly suppressed by high kinematic thresholds in cosmic rays (CRs) interactions, making them excellent probes for indirect detection of dark matter.
Experiments like the Alpha Magnetic Spectrometer (AMS-02)~\cite{AMS2015prl}, the General Antiparticle Spectrometer (GAPS)~\cite{GAPS2015}, 
and the Balloon-borne Experiment with a Superconducting Spectrometer (BESS)~\cite{BESS2005prl,BESS:2024yma} have been hunting for their signals for many years. Although the BESS program found no candidates, it has reported an upper limit for the differential flux of cosmic-ray antideuterons of ${\rm 1.9\times10^{-4} \rm (m^2\, s\, sr\, GeV/n)^{-1}}$, at the 95\% confidence level, in the range of 0.17{-}1.15 GeV/n~\cite{BESS2005prl} at the upper atmosphere (about 38 km altitude in this work). The BESS-Polar II program recently reported a new upper limit on the antideuteron flux of ${\rm 6.7\times10^{-5} \rm (m^2\, s\, sr\, GeV/n)^{-1}}$~\cite{BESS:2024yma} at 95\% confidence level in an energy range from 0.163 to 1.100 GeV/n. The AMS-02 program has measured the antiproton flux~\cite{AMS2016antiproton,AMS:2021nhj}, and recently detected 7 candidate events of antideuterons~\cite{ams-antid}. Some new calculations, such as Ref.~\cite{ Korwar:2024ofe}, indicate that the AMS-02 and GAPS are capable to observe antideuteron. The GAPS experiment optimized specifically for low-energy cosmic-ray antinuclei, will provide a sensitivity to antideuterons that was estimated to be ${\rm 2.0\times10^{-6} \rm (m^2\, s\, sr\, GeV/n)^{-1}}$~\cite{GAPS2015,GAPS2023}. To date, although certain experimental analyses have documented exceedingly rare occurrences of antinuclei~\cite{ALICE:2022zuz,MRasa:2024zse,DAngelo:2024vyn,ALICE:2022pbb,Shukla:2020bql,Chen:2018tnh} among billions of samples, the origin of these particles remains highly uncertain~\cite{Ting:2320166}. Theoretical researches on antideuterons, especially the background calculation of cosmic-ray antideuterons, are thus of great importance for experimental exploration. 

The dominant background of light antideuterons mainly comes from secondary particles when primary cosmic rays (CRs) collide with the interstellar medium (ISM) or Earth's upper atmosphere. Additionally, low-energy antideuterons can be enhanced through the antideuterons' non-annihilating inelastic interactions with the medium, which called as the tertiary production and lead to a migration of antideuterons from the high-energy part to the low-energy part of the spectrum~\cite{083012Duperray,FDonato2001teritary}.
For the near-Earth detection of antideuterons from dark matter annihilation, it is crucial to accurately assess their background contributions from astrophysical and hadronic processes~\cite{summarize2020JCAP,083012Duperray}. In particular, both secondary Galactic and atmospheric antideuterons can be produced through the same nuclear reaction mechanism in high energy nuclear collisions involving incident cosmic rays (mainly $protons$) colliding with either ISM nuclei (mainly $Hydrogen$) in the Galaxy or atmospheric nuclei (mainly $Nitrogen$-N, $Oxygen$-O, $Carbon$-C ) in the Earth's atmosphere \cite{Huang2003antiproton}.  In the atmosphere medium, the secondary antideuteron flux mainly originates from the $p$ + N, $p$ + O and $p$ + C collisions. Typically, the production of secondary antideuteron is described by the coalescence model, which relies on a coalescence factor that is taken from experimental  measurements~\cite{083012Duperray,FDonato2001teritary,Largescales2020,506FDonato2008,Ibarra2013JCAP,Kachelriess:2020uoh}.   
This can be improved by Monte Carlo event generators which take into account two-particle momentum correlations and the coalescence process is imposed on an event-by-event basis~\cite{Shao:2022eyd}. Moreover, atmospheric antideuterons can scatter upward into space,  influencing the distribution of low-energy antideuterons observed in space. This upward scattering necessitates the use of techniques such as geomagnetic backtracing~\cite{GAPS2015,ams-antid} to trace the origin of these particles and distinguish them from astrophysical sources. Consequently, it is crucial to take into account the tertiary antideuteron background when analyzing data from these space-based and balloon-borne experiments~\cite{summarize2016Aramaki}.

In the study, we calculated the production of antideuteron near the Earth's upper atmosphere with the full phase-space distribution of antiproton and antineutron in $p$ + A collisions generated by a multiphase transport  (AMPT) model~\cite{Lin:2004en}. The fluxes of these antideuterons~\cite{Huang2003antiproton,Huang2007,atmosphere2005PLB,atmosphere1996proton,simon1998,LBMmodel1992,LBM2008} in atmospheric propagation are determined using the leaky box model \cite{LBMmodel1992} with the inclusion of tertiary component to describe the evolution process of the atmospheric antideuteron.

\section{Antideuteron production and propagation}
\label{method}
\subsection{Antideuteron production from collisions of cosmic rays and the atmosphere medium}
Antideuteron production in upper atmosphere can be calculated by the AMPT model~\cite{Lin:2004en, Lin:2021mdn} coupled with a dynamical coalescence~\cite{Chen2006COAL,Sun:2016rev,Sun:2015jta,Pu:2018eei,Pu:2024kfh} model. 
The AMPT model, which has two versions (Default and String melting) and consists of four main components (initial conditions, partonic interactions, hadronization of partonic matter, and hadronic interactions), was employed to obtain the full phase-space information of antinucleons. In the AMPT model, the initial conditions are obtained from the HIJING model \cite{HIJING1991,HIJING1994}, which provides spatial and momentum distributions of minijet partons and soft strings. In the string melting mechanism, both excited strings and minijet partons are transformed into partons. Zhang's parton cascade model \cite{Zhang1997} was used to simulate the strong interactions among partons. A simple quark coalescence model is employed to describe the conversion of these partons to hadrons. A relativistic transport (ART) model~\cite{Libaoan1995} was used to simulate interactions among the hadrons and corresponding inverse reactions, as well as resonance decays.  The AMPT model, which consists of these four parts, has been widely used to simulate the evolution of dense matter produced in high energy nuclear collisions \cite{Lin:2004en, Lin:2021mdn}. Specifically, the string melting version of the AMPT model effectively describes anisotropic flows and particle correlations in collisions of $pp$, $pA$ or $AA$ systems at RHIC and LHC energies. In this study, the string melting version was employed to simulate the collision mechanism between the CRs and the main component of Earth's atmosphere to provide pahase-space information of antinucleons.       

To describe the antideuteron production, we employ a coalescence model~\cite{Gyulassy:1982pe,Mattiello:1996gq,Scheibl:1998tk,Mrowczynski:1992gc,Chen:2003qj,Chen:2003ava,Chen2006COAL,Steinheimer:2012tb,Sun:2018mqq,Sun:2022rjh,Sun:2022xjr} which has been successfully in describing the light nuclei~\cite{Sun:2016rev,Sun:2015jta,Sun:2015ulc,Gomez-Coral:2018yuk} production in relativistic heavy-ion collisions.  
In this model, the formation probability of an antideuteron from an antiproton and antineutron pair is given by the Wigner function  of the deuteron internal wave function, 
\begin{eqnarray}\label{wigner function}
\begin{aligned}
\rho_{\bar{d}}^{W}({r},{k}) =8\ \exp (-\frac{x^{2}}{\sigma_{\bar{d}}^{2}}-\sigma_{\bar{d}}^{2}{k}^{2}), 
\end{aligned}
\end{eqnarray}
where $k=(k_{1}-k_{2})/\sqrt{2}$, $x=(x_{1}-x_{2})/\sqrt{2}$, and $\sigma_{\bar{d}}=2/\sqrt{3}\ {r_{d}}$ with $r_{d}=1.96$ fm being its root-means-quare radius. Here, ${\bf x}_1$ and ${\bf x}_2$ are the spatial coordinates and ${\bf k}_1$ and ${\bf k}_2$ are the spatial momenta of the two nucleons in their rest frame at equal time, and they are obtained from propagating the nucleon with an earlier freeze-out time to the time of the later freeze-out nucleon. 

\subsection{Antideuteron propagation in the atmosphere}
\label{propagation}
The propagation of the produced antideuterons can be described by a leaky box model (LBM) ~\cite{LBMmodel1992,LBM2008} which has been successfully employed in previous studies~\cite{simon1998,083012Duperray}. 
The flux $\Phi_{\bar{d}}$ of the antideuteron at kinetic energy $T_{\bar{d}}$ is given by\cite{083012Duperray}:
\begin{eqnarray}\label{fluxcalculation}
{\Phi}(T_{\bar{d}})=\frac{\lambda_{esc}(T_{\bar{d}})\lambda_{int}(T_{\bar{d}})}{{\varrho}[\lambda_{esc}(T_{\bar{d}})+\lambda_{int}(T_{\bar{d}})]}\nonumber\\
\times\frac{1}{4\pi}[Q_{\bar{d}}^{sec}(T_{\bar{d}})+Q_{\bar{d}}^{ter}(T_{\bar{d}})]\,,
\end{eqnarray}
where the mean free path against inelastic interactions is ${\lambda_{int}(T_{\bar{d}})=\langle{m_{air}}\rangle/\langle{\sigma_{R}^{{\bar{d}+atm}}}(T_{\bar{d}})}\rangle$ and the $Q_{\bar{d}}^{sec}(T_{\bar{d}})$ and $Q_{\bar{d}}^{ter}(T_{\bar{d}})$ are the secondary and tertiary source terms, respectively. The average cross section on atmospheric gas is denoted by $\sigma_{R}^{{\bar{d}+atm}}=105A^{2/3}$ mb obtained from a parameterized formula in Refs.~\cite{083012Duperray,Binon1970CS}. In the LBM, $\lambda_{esc}$ is often parameterized based on the particle's rigidity (momentum per charge) or velocity. Different studies may adopt various parameterizations of $\lambda_{esc}$ to account for different propagation scenarios and observational constraints.
We used the LBM escape length $\lambda_{esc}(T_{\bar{d}}){=}11.8 \ \rm {g} /\rm {{cm}^{2}}$, which is taken from Refs.~\cite{Huang2003antiproton,Huang2007,simon1998}, and the average density $\varrho=2.28\times10^{-24}\ \rm {g} /\rm {{cm}^{3}}$, as suggested in Refs.~\cite{083012Duperray,atmosphere1996proton,simon1998}. The quantity $\langle m_{air}\rangle $ stands for the mean mass of the target air which has a value of $\langle m_{air}\rangle{=} 14.58\ amu\ (Nitrogen,\ Oxygen,\ Carbon)$. The atmosphere composition number density is simply set as ${\rm N}:{\rm O}:{\rm C} \ {=} \ {1} \ {:} \ {0.27}\ {:} \ {\rm 5.1\times10^{-4}} \ \rm{cm^{-3}}$ for their different proportions in the atmosphere. Actually, these numbers as well as the average density are not perfectly known, which results in a peculiar parametrization for $\lambda_{esc}(T_{\bar{d}})$~\cite{083012Duperray,simon1998}. However, an alternative choice for these quantities would not significantly affect the shape of the antideuteron flux distribution~\cite{083012Duperray}.

The antideuteron flux in the Galactic ISM are calculated by many researchers~\cite{083012Duperray,Kachelriess2020jcap,Laura2022prd,506FDonato2008}. Here,  the antideuterons produced from ISM are not included in our calculations. As described in Ref.~\cite{083012Duperray}, the secondary and tertiary atmospheric antideuteron flux was calculated using the same method as the calculation of antideuteron flux in the Galactic ISM with taking into account some specific processes and mechanism, such as, the specific nuclear reactions $p$ + N, $p$ + O and $p$ + C collisions and their secondary contribution to the flux. Meanwhile, the main composition and proportion of the atmosphere should also be considered. In the LBM model, the secondary source term $Q_{\bar{d}}^{sec}$ and the tertiary source term $Q_{\bar{d}}^{ter}$ in Eq. (\ref{fluxcalculation}) play curcial roles in the overall production of antideuteron from CRs that interact with the atmosphere. For the $\bar{d}$ flux, the secondary source term is given as~\cite{083012Duperray,simon1998}:
\begin{eqnarray}\label{Qsec}
\begin{aligned}
{Q}^{sec}(T_{\bar{d}})=& \sum _{i=\rm CRs}^{p}     \sum _{\rm j=\rm atm}^{\rm N,\rm O,\rm C} \ 4\pi n_{j} \int_{T_{min}}^{\infty} \frac{d\sigma^{i+j}(T_{\bar{d}}, T_{i})}{dT_{\bar{d}}}\\ 
&\times{\Phi_{i}(T_{i})dT_{i}},
\end{aligned}
\end{eqnarray}
where ${n}_{j}$ represents the number density of the atmospheric gas in $\rm {cm}^{-3}$, such as: ${n}_{\rm N}=1\ \rm {cm}^{-3}$, ${n}_{\rm O}=0.27\ \rm {cm}^{-3}$ and ${n}_{\rm C}={\rm 5.1\times10^{-4}} \ \rm{cm^{-3}}$ as mentioned earlier, while $ T_{\bar{d}}=(E_{\bar{d}}-m_{\bar{d}})/n$ denotes the antideuteron kinetic energy per nucleon, where n is the number of nucleon. The differential cross section $\sigma^{i+j}$ for the production of antideuterons at different energies ${T}_{i}$ follows~\cite{Kachelriess:2020uoh}: 
\begin{eqnarray}
\begin{aligned}
\frac{d\sigma^{i+j}(T_{\bar{d}}, T_{i})}{dT_{\bar{d}}}={\sigma_{ij}}\frac{{d}{N}_{\bar{d}}(T_{\bar{d}}, T_{i})}{dT_{\bar{d}}},
\end{aligned}
\end{eqnarray}
where ${\sigma_{ij}}$ is the total inelastic cross section of antideuteron for the proton with kinetic energy $T_{i}$ reacting with a ``fixed-target'' $j$ in the atmosphere. The ${\sigma_{ij}}$ can be calculated with antideuteron yields, which were obtained from ${p}+{\rm N/O/C}$ collisions by using the AMPT model coupled with a dynamical coalescence mode. The ${{d}{N}_{\bar{d}}(T_{\bar{d}},T_{i})}/{dT_{\bar{d}}}$ represents the kinetic energy distribution of antideuteron yields. The $\Phi_{i}(T_{i})$ is the flux of cosmic ray protons, which can be obtained by parameterizing the primary cosmic ray flux based on several space-based experimental measurements~\cite{AMS2015prl,DAMPE2019,CREAM-III,NUCLEON}. Following Ref.~\cite{Kachelriess:2020uoh}, a well-fitting parameterized formulae for the primary proton cosmic ray flux is given as:
\begin{eqnarray}
{\Phi}({T})={a}{T^{-{\gamma}}}(\frac{T}{T+b})^{c} \prod_{i=1}^{N} f(T_{bi},{\Delta}{\gamma_{i}},s),
\end{eqnarray}
where 
\begin{eqnarray}
f(T_{b},{\Delta}{\gamma},s)=[1+(\frac{T}{T_{b}})^{s}]^{{\Delta}{\gamma}/{s}}.
\end{eqnarray}
The parameters are taken for fits of the proton flux following Ref.~\cite{Kachelriess:2020uoh} with $N=2, a=26714$ m$^2$s sr/(GeV/n), $b=0.49$ GeV/n, $c=6.81$, $\gamma=2.88$, $T_{b1}=343$ GeV/n, $T_{b2}=19503$ GeV/n, $\Delta\gamma_1 =0.265$, $\Delta\gamma_2 =-0.264$, and $s=5$. 

Once antideuterons are formed, their inelastic interactions could result in a reduction of their flux, which can be described as the tertiary contribution. The tertiary source term $Q^{ter}$ was emphasized to describe the process involving the non-annihilating interaction of antideuterons with the main components of the upper atmosphere~\cite{FDonato2001teritary,506FDonato2008}. The tertiary contribution does not mean generating new antideuterons. It simply indicates that the flux of antideuteron at energy ${T}$ must account for the added redistribution of those at energies ${T}^{'}>{T}$, subtracting the flux of antideuterons redistributed to lower energies. This contribution can arise in various ways. Such as, antideuterons may undergo elastic scatterings process, but the cross section for these interactions is so minimal that the resulting energy loss of antideuterons is usually negligible. During elastic scatterings, antideuterons can exist without any change in energy. Additionally, antideuterons may undergo annihilate during their propagation, especially at low energies. However, in our current work, we did not employ an empirical approach to calculate the annihilation cross section for antideuterons according to Refs.~\cite{083012Duperray,506FDonato2008}, in this work we adopt the tertiary source term following~\cite{simon1998}:
\begin{eqnarray}\label{Qter}
\begin{aligned}
{Q}^{ter}(T_{\bar{d}})=&4{\pi}{n_j}\int_{T_{\bar{d}}}^{\infty}[\frac{{\sigma_{\rm inel}^{\bar{d}+atm{\rightarrow}\bar{d}X}}(T_{\bar{d}}^{'})}{{\sigma_{\rm inel}^{\bar{d}+p{\rightarrow}\bar{d}X}(T_{\bar{d}}^{'})}}]\\
&\times \frac{d\sigma^{\bar{d}+p{\rightarrow}\bar{d}X}}{dT_{\bar{d}}} 
(T_{\bar{d}}^{'}, T_{{\bar{d}}}) {\Phi}_{\bar{d}}({T_{\bar{d}}}^{'})dT_{\bar{d}}^{'} \, ,
\end{aligned}
\end{eqnarray}
where ${{\sigma_{\rm inel}^{\bar{d}+atm{\rightarrow}\bar{d}X}}(T_{\bar{d}}^{'})}$ represents the non-annihilating inelastic rescattering cross section between the produced antideuteron and the atmospheric medium, which can be derived from a parametric form based on Glauber calculations~\cite{Laura2022prd,PLB2011param,ALICE2020prl}. It can be parametrized as a function of atomic mass number of the target nucleus as~\cite{PLB2011param}:
\begin{eqnarray}
\sigma^{ h A}_{\text {inel}}= \pi R_{A}^{2} \ln \left[1+\frac{A \sigma^{hN}_{tot}}{ \pi R_{A}^{2}}\right],
\end{eqnarray}
where the total cross section \(\sigma_{\text{tot}}^{h N}\) of a hadron \text{h} (h= $\bar{p}$, $\bar{d}$, $\mbox{}^{3}\overline{He}$, $\mbox{}^{4}\overline{He}$) interacting with a
nucleon N is estimated with Glauber calculations~\cite{PLB2011param}. A is the
atomic number of the target nucleus with radius $R_{A}$.
The ${{\sigma_{\rm inel}^{\bar{d}+p{\rightarrow}\bar{d}X}(T_{\bar{d}}^{'})}}$ is the inelastic rescattering cross section for the antideuteron reacting with proton. Since no specific experiment was undertaken for this process, one can assume the antideuteron inelastic cross section as simply twice the size of the corresponding antiproton inelastic cross section~\cite{083012Duperray,506FDonato2008},
\begin{eqnarray}
{{\sigma_{\rm inel}^{\bar{d}+p{\rightarrow}\bar{d}X}(T_{\bar{d}})}}=2{{\sigma_{\rm inel}^{\bar{p}+p{\rightarrow}\bar{p}X}(T_{\bar{p}})}},
\end{eqnarray}
where ${{\sigma_{\rm inel}^{\bar{p}+p{\rightarrow}\bar{p}X}(T_{\bar{p}})}}$ was derived as~\cite{Tan1983,Laura2022prd,Tan1983}:

\begin{eqnarray}
\begin{aligned}
\sigma_{\text {inel }}^{\bar{p}+p{\rightarrow}\bar{p}X}\left(T_{\overline{\mathrm{p}}}\right)= & 24.7\left(1+0.584 \left(T_{\overline{\mathrm{p}}}\right)-0.115\right. \\& \left.+0.856 \left(T_{\overline{\mathrm{p}}}\right)-0.566\right) \text { mbarn }.\end{aligned}
\end{eqnarray}
The differential cross section in Eq.~(\ref{Qter}) indicates inelastic non-annihilating (NAR) process with incident energy $T_{\bar{d}}^{'}$ in low energy can be given by \cite{FDonato2001teritary}:
\begin{eqnarray}
\begin{aligned}
\frac{d\sigma^{\bar{d}+p{\rightarrow}\bar{d}X}}{dT_{\bar{d}}}(T_{\bar{d}}^{'}{\rightarrow}T_{\bar{d}})=\frac{{\sigma_{\rm NAR}^{\bar{d}+p{\rightarrow}\bar{d}X}}}{T_{\bar{d}}^{'}},
\end{aligned}
\end{eqnarray}
where ${\sigma_{\rm NAR}^{\bar{d}+p{\rightarrow}\bar{d}X}}$ denotes the total inelastic scattering cross section for $\bar{d}\,+p {\rightarrow} \bar{d}\,+X$, as detailed in Refs.~\cite{083012Duperray,506FDonato2008}. The tertiary contribution was treated as a corrective term and handled iterative. The ${\Phi}_{\bar{d}}(T_{\bar{d}}^{'})$ in the integral was calculated by using Eq. (\ref{fluxcalculation}) without $Q^{ter}$ contribution, while ${\Phi}_{\bar{d}}(T_{\bar{d}})$ necessitated an iterative approach for determination~\cite{506FDonato2008}.

\section{results and discussion}
\label{RAD}
\subsection{Model validation for antideuteron production in $ p +A\rightarrow \bar{d}+X$ collisions}
\begin{figure}[htbp]
\centering
\includegraphics[width=8.5cm]{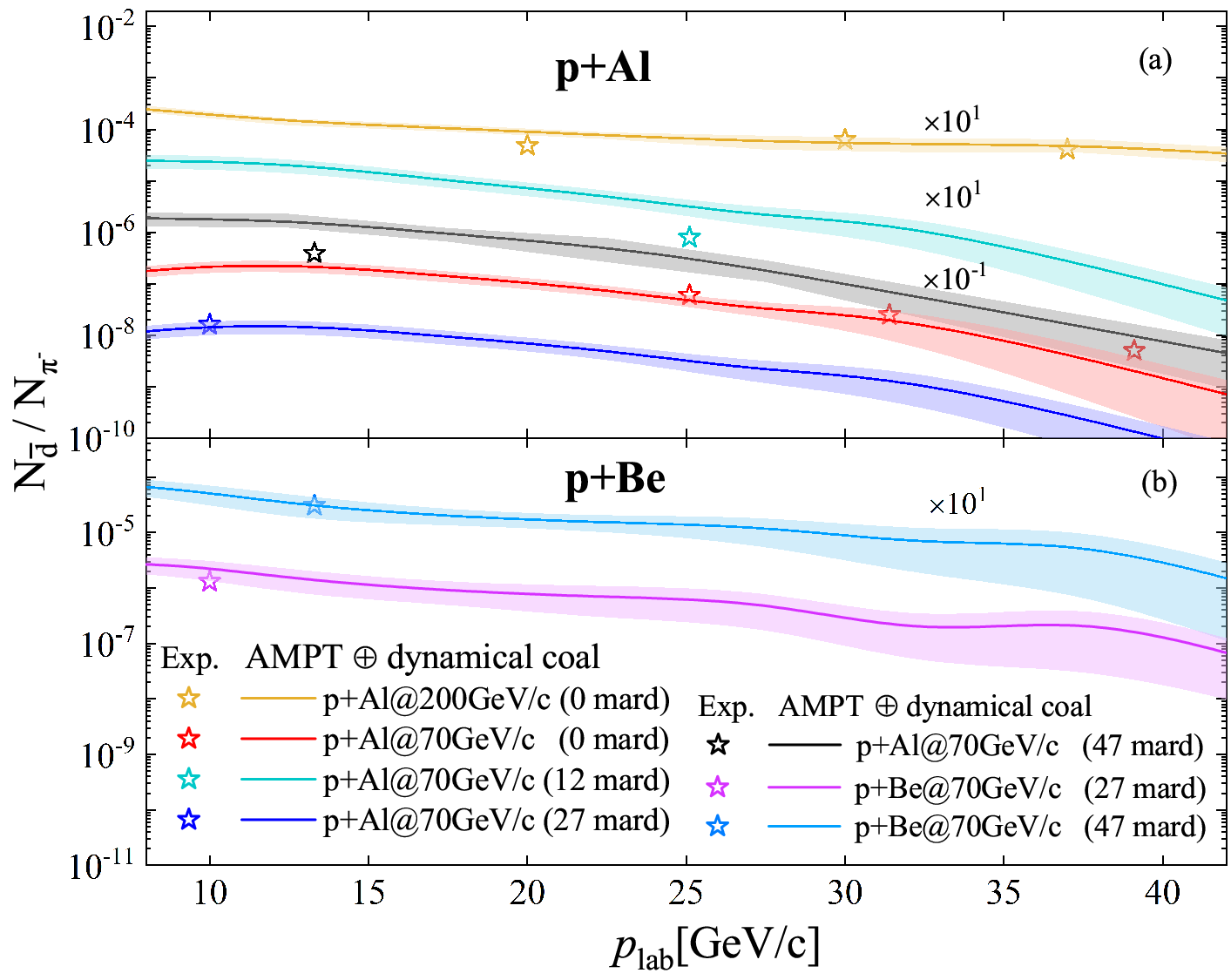}
\caption{\label{70200exp}(Color online) 
Results of antideuteron to pion yield ratios for this work (lines) comparing with the experimental data (stars) from $p$\,+A collisions at 70 GeV/c or 200 GeV/c as reported in Refs.~\cite{200GeV,70-1,70-2,70-3,70-4}. The top panel displays data measured at a range of angles: $0, 12, 27, 47$ mrad for $p$+Al collision, while the bottom panel exhibits data measured in $27, 47$ mrad at 70 GeV/$c$ for $p$+Be collisions. The shaded areas are the corresponding  uncertainties.
} 
\end{figure}

To validate our model and account for significant nuclear reactions between cosmic rays and atmospheric constituents, based on existing experimental measurements, the antideuteron production in $p$ + A collisions were calculated using the AMPT model coupled with a dynamical coalescence model. In this work, the ratios of antideuteron to pion yields $N_{\bar{d}\,}/N_{\pi^-}$ that produced from model calculation with 0.1 million simulations of $p$ + Al and $p$ + Be collisions at 200 GeV/$c$ and 70 GeV/$c$ respectively, are compared with that in experiments. Figure~\ref{70200exp} displays the comparison of $N_{\bar{d}\,}/N_{\pi^-}$ in $p$ + Al and $p$ + Be collisions~\cite{200GeV,70-1,70-2,70-3,70-4} between model calculation and experimental data for various collision energies. It is seen that our model calculation reproduces the data well from $p_{\rm lab}=10$ GeV/$c$ to 40 GeV/$c$, suggesting the approach is suitable for describing antideuteron production in high-energy $p$A collisions.

\subsection{Atmospheric antideuteron production}
Antideuterons can be formed by cosmic ray interactions with the atmosphere medium when the threshold energy is exceeded. Since the abundance of cosmic ray helium is significantly smaller compared to proton, we neglected the antideuteron contributions produced by helium reacting with atmosphere. The different contributions for $\frac{d\sigma^{i+j}}{dT_{\bar{d}}}$ caused by the breaks in the primary spectra are negligible below 10 GeV/n and small at higher energies according to the Ref.~\cite{Kachelriess:2020uoh}. So in the current work, to account for both the continuous of the cosmic ray proton spectrum and the essential threshold required for antideuteron formation, we parameterize the continuous cosmic ray protons spectrum within the energy range from 10 GeV/n to $5\times 10^{4}$ GeV/n~\cite{Kachelriess:2020uoh}. Upon determining the total integral of the spectrum, we divided the continuous proton spectrum into 9 energy ranges based on the fraction of the total integral for incident proton energy as shown in TABLE~\ref{TableEnergyBins}. The median value $T_p$ in each enrgy range was chosen as the incoming energy for simulating antideuteron production using the AMPT model coupled with the dynamical coalescence model. It is equivalent to that the proton energy spectra are split into 9 bins with various incident energies~\cite{Largescales2020} considering the weighting of different incident proton energy bins in the parameterized primary proton cosmic ray flux spectra~\cite{Kachelriess:2020uoh} which are shown in TABLE~\ref{TableEnergyBins}. In each energy channel, we use $T_{p}$ as the input energy to simulate the production of antideuteron multiplying the differential production distribution by its weight. We 
have simulated 0.5 million events for each channel of $i+j$ collisions ($i \in \{{p}\}$ and $j$ for $j \in $ \{{N, O, C}\}) in every incident energy. These results of kinetic energy distribution of antideuteron yields $dN_{\bar{d}}/dT_{\bar{d}}$ from model calculations serve as inputs for the flux calculation according to Eqs.~(\ref{fluxcalculation}) and (\ref{Qsec}). The cross sections of secondary antideuteron $\sigma_{\bar{d}}^{p+\text{N/O/C}}$ from model calculations for $p$ + N/O/C in different energies are summarized in TABLE~\ref{TableEnergyBins}.

\begin{figure}[htbp]
	\flushleft
	\includegraphics
	[width=8.5cm]{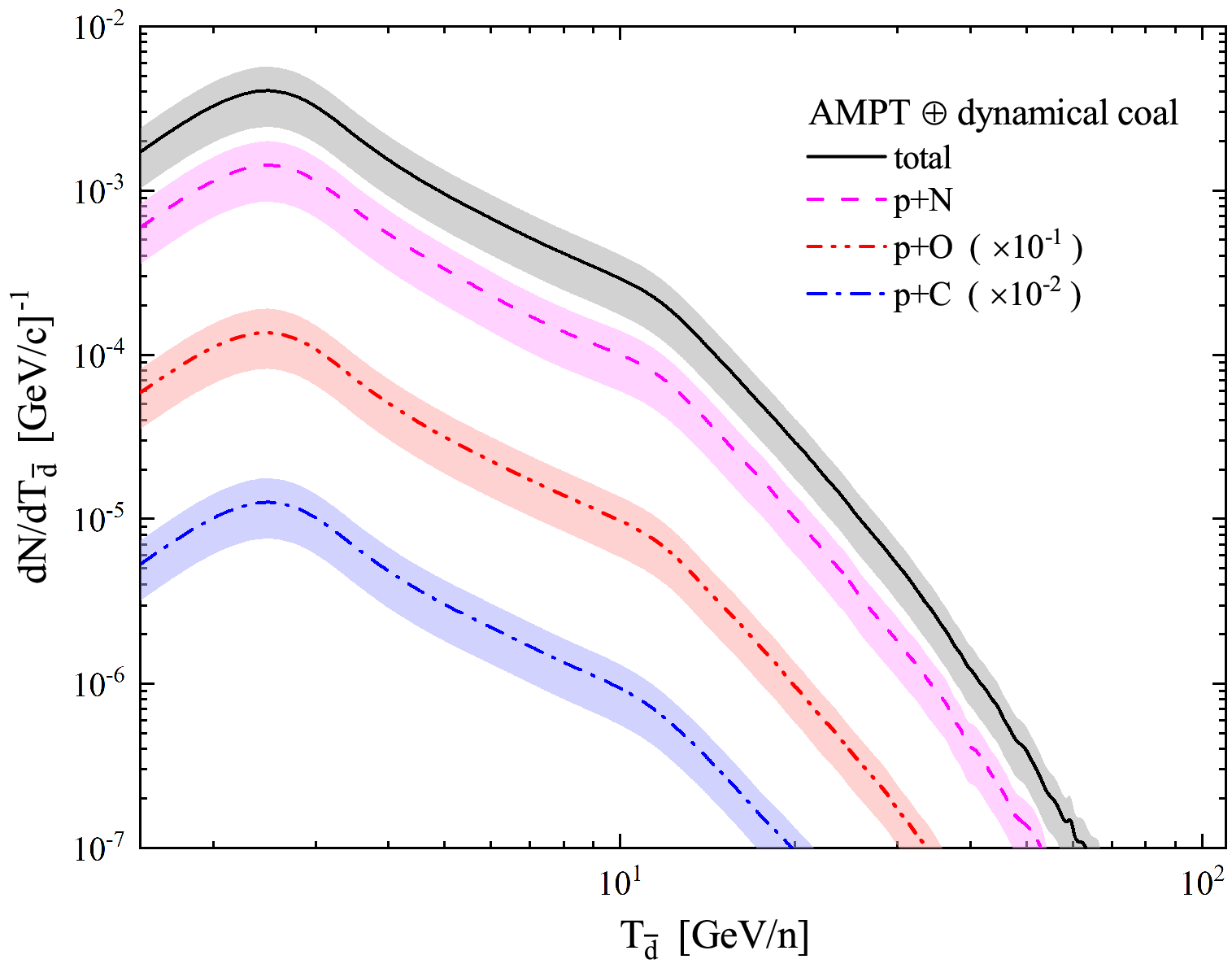}
	\caption{\label{pNproduction.eps} (Color online) 
		The kinetic energy distributions of secondary antideuterons for the  $p$ + N, $p$ + O and $p$ + C reactions are calculated using the AMPT model coupled with a dynamical coalescence model. The shaded areas are the corresponding statistical errors. 
	}
\end{figure}

\begin{table*}[htbp]
\centering
\caption{\label{TableEnergyBins} Incident proton energy, upper ($T_{Up}$) and lower ($T_{Low}$) limit values of the energy bins, converted collision energies in the center-of-mass, the proton flux and corresponding weight. The cross sections with statistical errors of secondary antideuteron from model calculations for $p$+N/O/C with corresponding energies.}
\begin{tabular}{p{1.5cm}p{1.5cm} p{1.5cm} p{1.2cm} p{3.8cm} p{1.5cm} p{1.8cm} p{1.8cm} p{1.8cm} }
\hline
$T_{p}$ &$T_{Low}$  &$T_{Up}$  &$\sqrt{s}$  &Flux &weight & $\sigma_{\bar{d}}^{p+N}$ &$\sigma_{\bar{d}}^{p+O}$ &$\sigma_{\bar{d}}^{p+C}$ \\
&  &   &  &   &  &  (stat.error) & (stat.error) &(stat.error)\\
\hline
(GeV/n) & (GeV/n) &(GeV/n) & (GeV)& [$\rm m^{-2}$$\rm s^{-1}$$\rm sr^{-1}$$(\rm GeV/n)^{-1}$] &($\%$) &(mb) &(mb) &(mb)\\
\hline \hline
20      & 10        & 30        & 6.1    & 4.056       & 85.40  & 2.130E-5 (6.641E-6)  & 1.810E-5 (4.395E-6) & 2.640E-5 (8.555E-6) \\
40      & 30        & 50        & 8.6    & 5.981E-1    & 8.83   & 9.110E-4 (3.413E-4) & 1.050E-3 (3.551E-4) & 9.380E-4 (3.0344E-4)\\
80      & 50        & 110       & 12.2   & 8.468E-2    & 4.42   & 7.760E-3 (1.860E-3) & 8.400E-3 (1.040E-3) & 7.990E-3 (1.590E-3)\\
200     & 110       & 290       & 19.3   & 6.220E-3    & 1.12   &  3.510E-2 (7.3E-3) & 3.881E-2 (7.430E-3) & 3.542E-2 (9.460E-3)\\
750     & 290       & 1210      & 37.3   & 1.718E-4    & 0.22   & 1.170E-1 (1.101E-2) & 1.267E-1 (2.080E-2) & 1.205E-1 (2.900E-2)\\
1900    & 1210      & 2590      & 59.4   & 1.514E-5    & 0.017  & 1.724E-1 (2.569E-2) & 1.830E-1 (3.446E-2) & 1.670E-1 (3.406E-2)\\
4700    & 2590      & 6810      & 93.5   & 1.419E-6    & 0.0056 & 2.541E-1 (4.733E-2) & 2.588E-1 (6.290E-2) & 2.268E-1 (5.341E-2)\\
12500   & 6810      & 18190     & 152.6  & 1.093E-7    & 0.00118& 3.521E-1 (7.331E-2) & 3.737E-1 (6.081E-2) & 3.411E-1 (9.040E-2)\\
34095   & 18190     & 50000     & 252    & 6.862E-9    & 0.00022& 5.084E-1 (1.047E-1) & 5.507E-1 (1.038E-1) & 4.633E-1 (1.099E-1)\\ 
\hline
\end{tabular}
\end{table*}
Figure~\ref{pNproduction.eps} illustrates the antideuteron production calculated using the AMPT model coupled with the coalescence model to simulate CRs ($p$) interacting with Earth's atmospheric constituents (N/O/C). The differential yield distribution with antideuteron momentum varies with reaction types and shows a trend of first increasing and then decreasing. Notably, the $p$ + N reaction significantly dominates the production, followed by the $p$ + O reaction which contributes slightly less, and the $p$ + C reaction with minimal contribution. This trend may be attributed to the selection of $n_{j}$ as discussed in section~\ref{propagation}. As a matter of fact, in the atmosphere of Earth, nitrogen comprises the majority of atmospheric gas, while oxygen constitutes approximately 25\% of the total, and carbon only accounts for 0.03\%. FIG.~\ref{pNproduction.eps} shows a concave shape between the peak around 3 GeV/n and the shoulder around 10 GeV/n. Because the primary proton spectrum~\cite{Kachelriess:2020uoh} has two peaks, the weighted sum of the antideuteron production distribution over 9 energy ranges leads to the observed concave shape between the peak around 3 GeV/n and the shoulder around 10 GeV/n.

\begin{figure}[htbp]
\flushleft
\includegraphics
 [width=8.5cm]{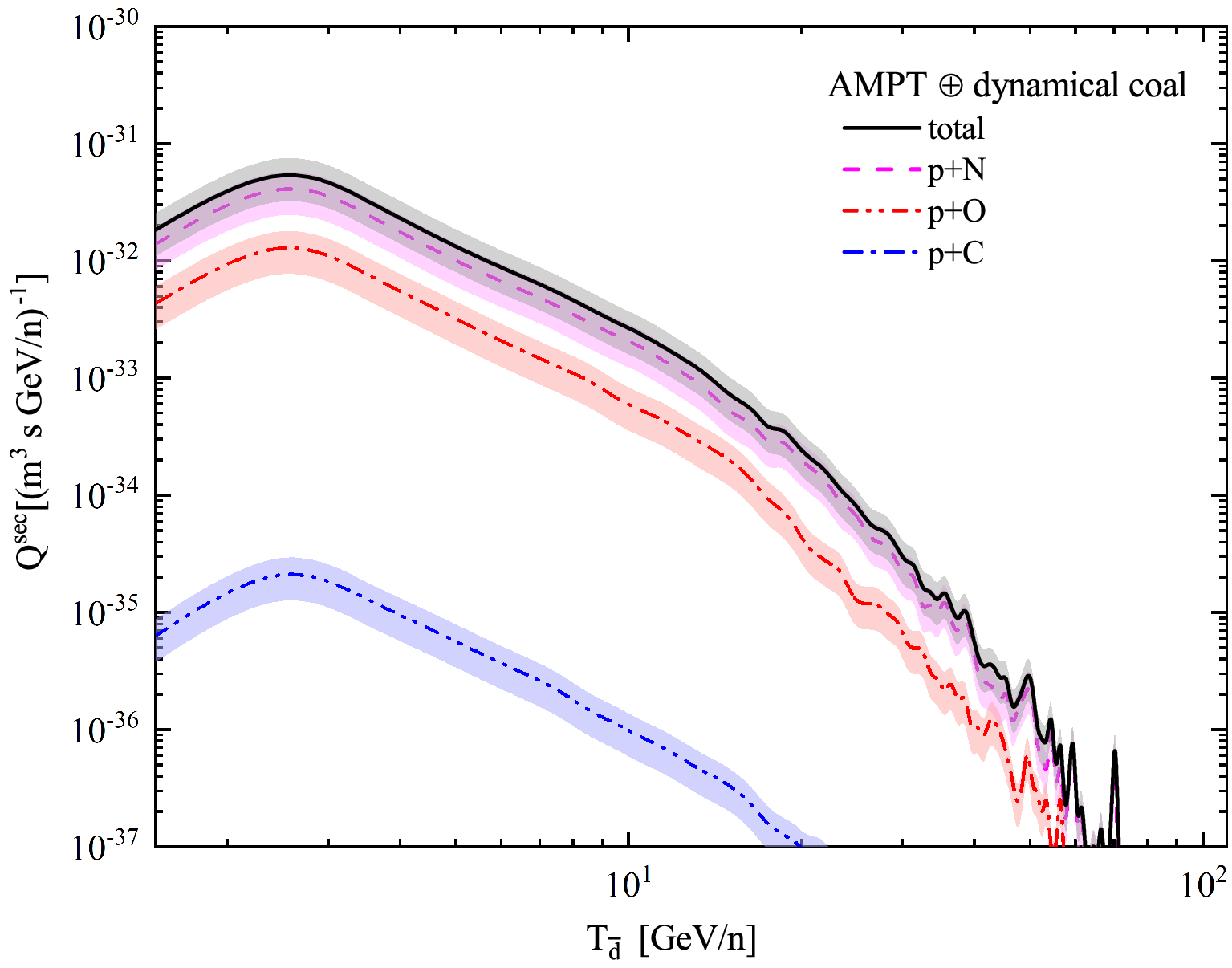}
 \caption{\label{Qsecsource.eps} (Color online) 
The distribution of antideuteron kinetic energy per nucleon as a function of $Q_{sec}$, defined in relation (\ref{Qsec}), encompasses contributions from various reaction types to the total secondary source term arsing from cosmic ray protons interacting with target particles in the Earth's atmosphere. The shaded areas are the corresponding  statistical errors.
}
\end{figure}

Figure~\ref{Qsecsource.eps} shows the diverse contributions of the aforementioned processes to the source term $Q_{sec}$ for different reaction types, as defined in Eq.~(\ref{Qsec}). Regarding the secondary source $Q_{sec}$ for antideuterons, the contributions from $p$ + N, $p$ + O and $p$ + C collisions exhibit a similar distribution, peaking around 2$-$3 GeV/n. This feature can be explained by the threshold effect~\cite{Antinucci:1972ib} that the $\bar{d}$ flux distribution in the low energy region is primarily influenced by the rapidly increasing $\bar{d}$ production cross section when the energy surpasses its production threshold~\cite{083012Duperray}. The high energy region decay of the distribution is determined by the rapidly declining flux of incident cosmic ray protons with increasing energy, combined with the natural decrease of the high energy production cross section. And at every incoming energy level, the $p$ + N reaction prominently contributes to the production of $Q_{sec}$, with the $p$ + O reaction providing a slightly less contribution to the $p$ + N reaction.          

\subsection{Atmospheric antideuteron flux calculation}
The propagation of antideuteron in the atmosphere can be predicted by a widely used LBM model including a tertiary contribution~\cite{simon1998,506FDonato2008,083012Duperray}. As discussed in the section~\ref{propagation}, a numerical method was employed to solve Eq.~(\ref{fluxcalculation})~\cite{simon1998,506FDonato2008,083012Duperray} for forecasting the flux of atmospheric antideuterons resulting from interactions between cosmic rays (predominantly protons) and the atmosphere.
In reality, CRs particles traverse a considerable portion of the Earth's atmosphere before being detected by experiments~\cite{AMS2015prl,GAPS2015,DAMPE2019,BESS2005prl,BESSII2021}. Following Refs.~\cite{083012Duperray,Huang2003antiproton}, the amount of material encountered by CRs during this journey can be comparable to that encountered during their traversal through the Galaxy. The same mechanism of produce the antideuteron flux is used through interactions of CRs particles with the atmosphere.

 \begin{figure}[htbp]
\flushright
\includegraphics
 [width=8.5cm]{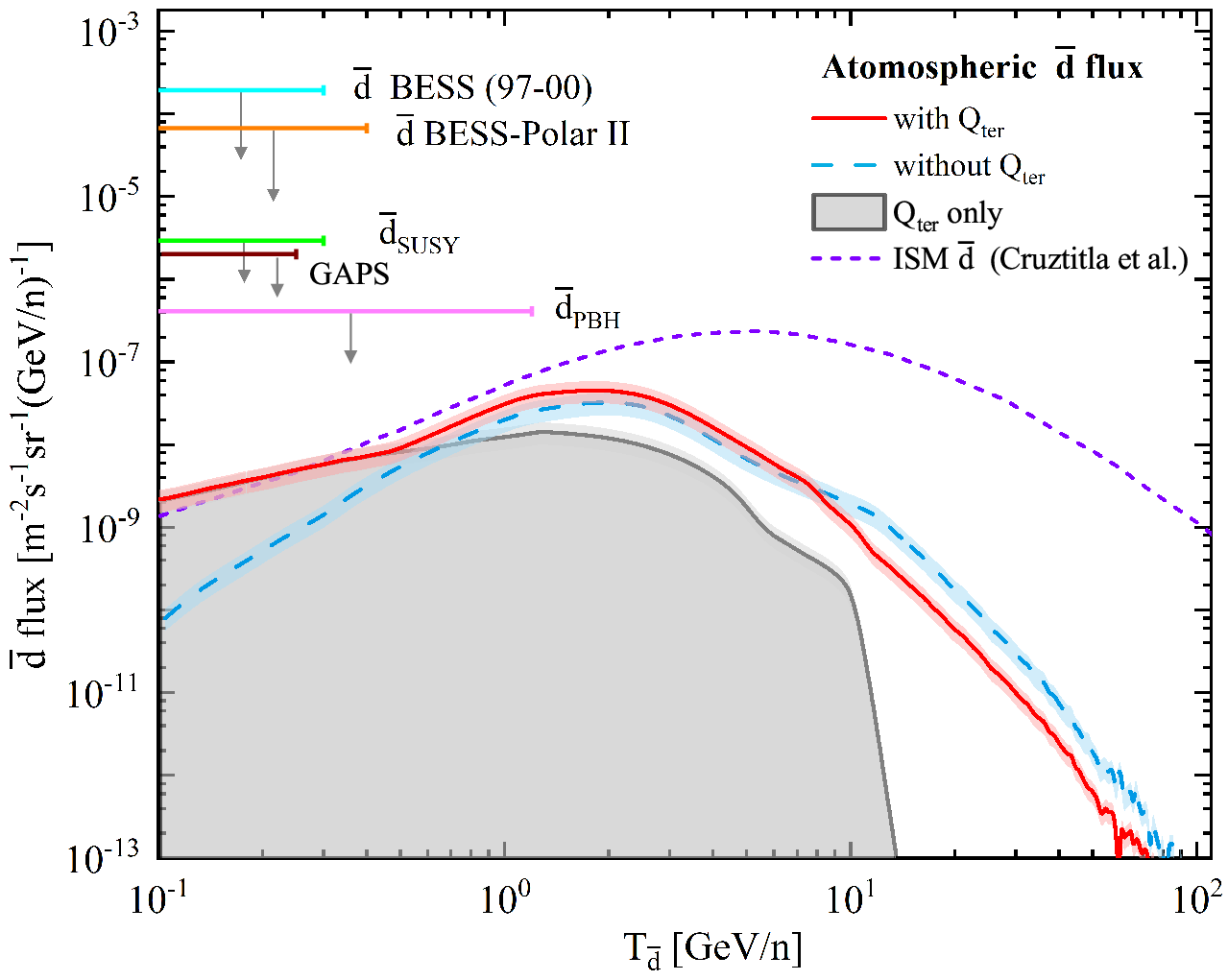}
 \caption{\label{total flux}(Color online) Calculated atmospheric $\bar{d}$ flux produced in the upper atmosphere (at an altitude of 38 \textit{km}) is compared with the upper limits observed by BESS 97-00 and BESS-Polar
\uppercase\expandafter{\romannumeral2} experiments. The red solid line represents the total $\bar{d}$ flux including $Q_{ter}$ contributions, while the blue dash line indicates the $\bar{d}$ flux excluding $Q_{ter}$ contribution. The red and blue shaded areas are the corresponding  uncertainties. The black line and shaded areas represent the $\bar{d}$ flux contributed solely by $Q_{ter}$. The violet short dash line represent the estimated ISM secondary $\bar{d}$ flux with GALPROP v.57 from Ref.~\cite{Cruztitla:2024nwr}. The horizontal line labeled ${\bar{d}}$ BESS(97-00) corresponds to the upper limit on $\bar{d}$ flux observed by BESS experiments from 1997 to 2000~\cite{BESS2005prl}. The horizontal line labeled ${\bar{d}}$ BESS-Polar
\uppercase\expandafter{\romannumeral2} corresponds to the updated upper limit on $\bar{d}$ flux observed by BESS experiments~\cite{BESS:2024yma}. The horizontal line labeled GAPS corresponds to the upper limit on $\bar{d}$ flux observed by GAPS experiments~\cite{GAPS2015}. Additionally, the horizontal line labeled $\bar{d}_{\rm SUSY}$ represents the upper limit on $\bar{d}$ flux from super-symmetric dark matter annihilation~\cite{Donato:1999gy}. The horizontal line labeled $\bar{d}_{\rm PBH}$ illustrates the upper limit on evaporative products of primordial black holes~\cite{083012Duperray,Barrau:2002mc}.        
}
\end{figure}    

According to Eqs.~(\ref{fluxcalculation}) to (\ref{Qter}), antideuterons are generated from collisions between the atmosphere and cosmic ray protons with energy ranging from 20 GeV/n to $5\times 10^{4}$ GeV/n. FIG.~\ref{total flux} presents the calculated atmospheric antideuteron flux derived from the LBM model framework and simulations conducted using the AMPT model coupled with the dynamical coalescence model, as well as some experimental limitation and theory estimates. The total antideuteron flux comprising secondary and tertiary components, is depicted in red solid line. The blue dash line represents the secondary flux excluding contributions from the tertiary source term. The secondary antideuteron flux can reach approximately $\rm 10^{-9}\sim10^{-8}\,m^{-2}\,s^{-1}\,sr^{-1}\,(GeV/n)^{-1}$ from 0.5 GeV/n to 1 GeV/n. However, when the kinetic energie of antideuterons exceeds about 10 GeV/n, the flux diminishes notably, potentially due to the limited amount of cosmic rays with sufficient high energy to produce antideuterons. Meanwhile, propagation effects can still exert a subtle influence, especially in redistributing the high energy spectrum of antideuterons.        
The black solid line and shaded area denote the contribution of the tertiary flux. Tertiary production arises from non-annihilating inelastic interactions of cosmic-ray antideuterons, resulting in a redistribution of antideuterons at low energys. This tertiary mechanism also determines the non-annihilating inelastic processes, previously disregarded due to the small binding energy of CRs antideuterons with nucleons or nuclear targets~\cite{Donato:1999gy}. It is interesting to see that the role of the tertiary component primarily mainfests in the low-energy range, altering the energy distribution and enhancing the secondary component. 
The solid red line represents the total antideuteron flux encompassing contributions from $Q_{sec}$ and $Q_{ter}$. Upon considering all contributions, the total antideuteron flux exhibits an enhancement in the low-energy region, up to a kinetic energy of approximately 8 GeV/n, compared to the scenario without the $Q_{ter}$ contributions. However, at higher kinetic energies, a reduction of the total flux is observed. This tendency can be traced to the generation and underlying mechanisms of the $Q_{ter}$ contributions as previously described. 
The antideuteron flux reaches a maximum value of about $\rm 5.24\times 10^{-8}\,m^{-2}\,s^{-1}\,sr^{-1}\,(GeV/n)^{-1}$ at a kinetic energy of about 2.5 GeV/n, which is comparable with results in Ref.~\cite{083012Duperray}. For comparison, the estimated ISM secondary $\bar{d}$ flux~\cite{Cruztitla:2024nwr} is also depicted in Fig.~\ref{total flux}. It is found that the atomospheric antideuteron dominates the antideuteron background at kinetic energies below $0.26 $ GeV/n, whereas the ISM antideuteron flux becomes the primary background at kinetic energies above $0.26$ GeV/n.

The total antideuteron flux calculated in this study is approximately two orders of magnitude lower than the upper limit for antideuteron flux from supersymmetric dark matter annihilation (indicated by the horizontal line labeled $\rm \bar{d}_{SUSY}$) provided in Ref.~\cite{Donato:1999gy}. This discrepancy can be attributed to the rapid drop in the kinetic energy distribution of predicted dark matter signals. The horizontal line labeled $\rm \bar{d}_{PBH}$ represents the upper limit on evaporative products of primordial black holes (PBH)~\cite{083012Duperray}. It is evident that the antideuteron flux we have calculated remains approximately an order of magnitude lower than the result predicted from PBH~\cite{083012Duperray}. Regarding the experimental detection of cosmic-ray antideuteron background, the BESS experiment has provided an upper limit on antideuterons at approximately 38 km above sea level, although no comic-ray antideuterons were observed. The horizontal line labeled $\bar{d}$ BESS (97-00) experiment represents the upper limit for the $\bar{d}$ flux. It is evident that the total antideuteron flux we calculated differs significantly from the BESS(97-00) limit\cite{BESS2005prl}, with a discrepancy of approximately three orders of magnitude. Despite BESS-Polar
\uppercase\expandafter{\romannumeral2} analyzing more than ten times the experimental data compared to previous BESS efforts, no candidate antideuterons were detected, resulting in a new upper limit (indicated by the horizontal line labeled $\bar{d}$ BESS-Polar
\uppercase\expandafter{\romannumeral2})~\cite{BESS:2024yma,BESS2021newlimit}. Notably, our calculations span several orders of magnitude in value and remain significantly below the upper limits of the BESS experiments~\cite{BESS2005prl,BESS:2024yma,BESS2021newlimit}, suggesting that the optimal strategy for antideuteron detection may lie in the lower kinetic energy range. The horizontal line labeled GAPS corresponds to the sensitivity for the GAPS antideuteron search, which is about 3 orders of magnitude higher than our calculation. The GAPS antideuteron search obtained using a Monte Carlo simulation indicates that GAPS has a strong potential to probe a wide variety of dark matter annihilation and decay models through antideuteron measurements~\cite{GAPS2015}.

\section{summary}
\label{summary}
The antideuteron background induced by cosmic rays near Earth's atmosphere have been calculated in this work.
With the primary proton flux obtained from experimental measurements, the interaction of primary cosmic rays with Earth's atmosphere medium is modeled by the AMPT model, which is then coupled with a dynamical coalescence model to calculate antideuteron production. Subsequently, a leaky box model framework with the inclusion of tertiary component (inelastic non-annihilating process) has been employed to describe the interaction and propagation of the produced antideuterons in the atmosphere medium. Within this hybrid approach, we estimate the atmospheric antideuterons fluxes near Earth. The obtained antideuteron flux is found to increase about 5 orders of magnitude for antideuteron kinetic energy per nucleon from $T=100$ GeV/n to 2 GeV/n, suggesting that the optimal window of balloon-borne experiments for antideuteron detection lie in the low kinetic energy range $T<10$ GeV/n. The obtained antideuteron flux is significantly below the upper limit observed by the BESS experiments, which provides a valuable reference for balloon-borne experiments like BESS-Polar\uppercase\expandafter{\romannumeral2} and GAPS, as well as space-based experiments like AMS-02. 
Besides, it is found that the atomospheric antideuterons dominate the antideuteron background at kinetic energies below $0.26 $ GeV/n, whereas the intersterllar medium antideuteron flux becomes the primary background at kinetic energies above $0.26$ GeV/n.

Furthermore, the methodology developed in the present study can be extended to calculate antideuteron production in the interstellar medium. In the present analysis of cosmic ray propagation, we have predominantly considered the cosmic ray proton component due to its prevalence. However, other components, such as helium, have not been included in the present calculations due to their relatively lower abundance. Future investigations could optimize these models by incorporating these additional cosmic ray species and more detailed description of the propagation process, thereby achieving a more comprehensive and accurate understanding of antideuteron fluxes in various astrophysical environments.


\begin{acknowledgments}
We thank Gang Guo, Jin-Hui Chen, Jiang-Lai Liu  Meng-Jiao Xiao, Yue-Lin Sming Tsai and Xiao Wang for helpful discussions.
This work was supported by the National Key Research and Development Project of China under Grant No.~2024YFA1612500, the National Natural Science Foundation of China under Grant No.~12105079, 12422509, No.~12375121, No.~12235010, the National SKA Program of China No.~2020SKA0120300, the Science and Technology Commission of Shanghai Municipality under Grant No.~23JC1402700, and the Natural Science Foundation of Henan province No.~242300422048.   The computations in this research were performed using the CFFF platform of Fudan University.
\end{acknowledgments}
\newpage

\end{document}